\documentclass{aa} \usepackage{graphicx} \usepackage{txfonts} \usepackage{lipsum} 
\usepackage{lscape} \usepackage{placeins} \usepackage{longtable} \usepackage{natbib}
\usepackage{hyperref}
\usepackage{xcolor} 
\bibpunct{(}{)}{;}{a}{}{,}

\makeatletter
\let\linenumbers\relax
\let\@linenumbers\relax

\makeatother

\begin{document} 
\title{Wide binaries without viable bound Newtonian orbits}
 \subtitle{} 
 \author{L. Pasquini\inst{1} \and R. Saglia\inst{2} \and F. Patat\inst{3} \and L. Berni\inst{1,9} \and D. Bossini\inst{4,5} \and L. Magrini\inst{1}  \and H. Ludwig\inst{6} \and M. T. Murphy\inst{7} \and J.R. de Medeiros\inst{8} \and J. Chanam\'e \inst{10} \fnmsep\thanks{Based on ESO-VLT observations, programmes 106.A-220(A),106.A-220(B) and 108-A-0192(A); PI Chanam\'e.} } 
 \institute{INAF, Osservatorio Astrofisico di Arcetri, Largo E. Fermi 5, 50100 Firenze Italy \email{luca.pasquini@inaf.it} 
 \and Max Planck Institute for Extraterrestrial Physics, Giessenbachstr. 85748 Garching, Germany 
 \and ESO, Karl Schwarzschild Strasse 2, 85478 Garching bei München, Germany 
 \and Dipartimento di Fisica e Astronomia, Università degli studi di Padova, Vicolo dell'Osservatorio 3, Padova, Italy 
 \and INAF, Osservatorio Astronomico di Padova, Vicolo dell'Osservatorio 5, Padova, Italy 
 \and Landessternwarte Zentrum f\"ur Astronomie der Universit\"at Heidelberg, K\"onigstuhl 12, 69117 Heidelberg, Germany
 \and Centre for Astrophysics and Supercomputing, Swinburne University of Technology, Hawthorn, Victoria 3122, Australia 
 \and Departamento de Fisica, Universidade Federal do Rio Grande do Norte, 59078-970 Natal, RN, Brazil 
 \and Dipartimento di Fisica e Astronomia, Università degli Studi di Firenze, Via Sansone 1, 50019, Sesto Fiorentino, Italy
 \and Instituto de Astrofísica, Pontificia Universidad Católica de Chile, Av. Vicuña Mackenna 4860, 782-0436 Macul, Santiago, Chile}
\date{Received September 30, 20XX} 
\abstract {Wide binaries offer a unique opportunity to test gravity in the low-acceleration regime, where deviations from Newtonian dynamics may appear. } 
{We used high-resolution VLT–ESPRESSO archival spectra to study 26 wide binaries with projected separations >13,000 AU. By combining precise radial velocities with {\em Gaia} proper motions and parallaxes, we tested whether these systems are consistent with Newtonian gravity in the low-acceleration regime.} 
{We used multiple radial-velocity measurements and stellar parameters to remove systems affected by unresolved triple or chance alignments as well as young systems. 
For the remaining binaries, we combined radial velocities (corrected for convective shift and gravitational redshift) with {\em Gaia} proper motions, parallaxes, and positions in a bid to find bound Newtonian orbital solutions.} 
{Of the 26 initial systems, 14 were discarded: 12 due to radial-velocity variability indicating unresolved close binaries, 1 that hosts a faint {\em Gaia} companion, and 1 that is too young. Of the remaining 12, 9 can be fitted with a bound orbital solution, while the velocity differences of the other 3 are  too large to be reconciled with any   bound Newtonian orbit.}
{For the three systems that cannot be fitted with a  bound  orbit, repeated radial-velocity observations allowed us to confidently exclude, with one possible exception, unresolved triple stellar companions or massive close-in planets as causes. Given their likely large 3D separations, these binaries may have been dynamically perturbed or disrupted by stellar encounters or Galactic tides, and may no longer be gravitationally bound. This highlights how utmost caution must be applied when studying wide binaries  as isolated systems.}
\maketitle 
 
\section{Introduction}
Wide binaries (WBs) are common stellar systems whose dynamics offer a unique opportunity to test gravity in the low-acceleration regime, in addition to informing stellar and exoplanet studies.
Following  {\it Gaia} data releases (DRs), they have been the subject of a considerable number of studies, particularly after the seminal {\it Gaia} catalogue was published by  \citet{Elbadry2021}.
In this letter we concentrate on the use of WBs as probes of gravity in the regime of low gravitational potential \citep{Hernandez+2012}.
Wide binaries experience gravitational attraction comparable to that in the outer regions of galaxies. Since in the solar neighbourhood dark matter is negligible  \citep{Soding+2025}, 
WBs are suitable probes for directly testing the Newtonian gravitational law in the low-acceleration regime. One limitation is the length of their orbital periods ($\sim 10^4-10^7$ yrs), which precludes us from observationally reconstructing the full orbits.
A series of studies took advantage of the exceptional {\it Gaia} astrometry to create large WB samples with precise  {\it Gaia} projected separations and transverse velocities \citep{Pittordis+2019, Banik+2024, Chae+2023, Chae+2024} but reached opposite conclusions on whether Newtonian gravity best reproduces the observations.
Using HARPS  radial velocities, \citet{Saglia+2025} have shown that high-quality radial velocities are essential to removing multiple systems from the samples (e.g.\ where a third body orbits one of the WB stars) and that combining the radial velocities with {\it Gaia} DR3 parallaxes and proper motions makes it possible to constrain the orbits of the binaries.
These authors analysed 32 WBs, finding that 31 of them could be explained with a viable bound Newtonian solution.
Radial velocities have also been used by \citet{Chae2025a}, who used {\em Gaia}'s best-quality radial velocities and a Bayesian approach to compare Newtonian and MOdified Newtonian Dynamics (MOND) solutions, finding that the latter are preferred for separations beyond 2 kAU. Similar conclusions were reached by \citet{Chae2025b}, who applied a Bayesian analysis to the HARPS sample published by \citet{Saglia+2025}.

In this study we tested whether WBs with large separations can be reproduced by bound Newtonian  orbits (`Keplerian' orbits hereafter), applying the approach illustrated in detail by \citet{Saglia+2025} to a set of archival VLT-ESPRESSO observations of WBs with projected separations larger than 13 kAU.
This sample (see Table \ref{table1}) complements the WBs studied by \citet{Saglia+2025}, which have maximum projected separations of 12.3 kAU.

\section{Sample and observations}
Data were retrieved from the  \href{https://archive.eso.org}{ESO Archive}.
The stars were observed between December 2020 and March 2022 with the ESPRESSO high-resolution spectrograph at the VLT \citep{Pepe+2021}.
The spectra are of excellent quality, with  a final signal-to-noise ratio (S/N) that varies between 170 and 400  around 570~nm in the combined exposures.
ESPRESSO covers the entire 380--790~nm wavelength range at a high resolution ($R \sim 140{,}000$) and is designed to achieve the highest radial-velocity precision \citep{Pepe+2021}.
In practice, the radial-velocity precision of these observations is limited to a few~m~s$^{-1}$ by the S/N of the spectra and by the stellar characteristics (e.g.\ a high rotation for some targets), fully adequate for the purposes of this study.

Our WBs were taken from the list of solar-type WBs constructed using the Tycho--\textit{Gaia} DR2 catalogues and were chosen to have projected separations ($s$) of > $13$ kAU  \citep[][]{Andrews+2017}.
The selected binaries all have very similar components, differing by less than 0.3~mag in brightness.
Several observations were obtained for each star (a minimum of three and a maximum of ten), for a total of 52 stars in 26 binary systems.
All stars are either on the main sequence or just leaving it.

\section{Methods}

We used the methodology presented in \citet{Saglia+2025}.
Multiple radial velocities, together with metallicity [Fe/H]\footnote{[Fe/H]=$(\log \frac{n_{\rm Fe}}{n_{\rm H}})_{\star} - (\log \frac{n_{\rm Fe}}{n_{\rm H}})_{\odot}$ }, were used to identify multiple systems and chance alignments (i.e.\ non-genuine WBs).

We first analysed the radial velocities. Similarly to HARPS, ESPRESSO radial velocities were determined by cross-correlating a digital mask with the observed spectrum.
The numbers of observations and masks used are given in Table~\ref{tab:stellar_par}.
The radial-velocity uncertainty was computed by the ESPRESSO data reduction pipeline \citep{Modigliani2020ASPC..527..667M}, and for these observations the error associated with a single measurement is 3--9~ms$^{-1}$ (see Table~\ref{tab:stellar_par}).
Out of the 26 initial binaries, 12 were discarded because at least one of the components of the system  exhibited significant radial-velocity variability. The list of these stars, along with the reasons for discarding them, is provided in Appendix~\ref{appendix}.

We also used {\it Gaia} DR3 to investigate the presence of faint stars in the fields that share the WB's parallax and proper motions.
BD~-20~655 has a faint (G = 17.18) nearby companion with the same parallax and proper motions; therefore, we excluded this pair (BD-20655 and BD-20658)  from the final sample.

The mean re-normalised unit weight error (RUWE) index for the 26 surviving stars is 1.03$\pm$0.14, indicating that they behave as single stars with reliable astrometric solutions. Only TYC 5346-457-1 has a  RUWE of 1.44, hinting at a possible unresolved multiplicity. Since the radial velocity is very stable (1 m\,s$^{-1}$) over seven observations, we kept this star in the sample. 

For the determination of stellar parameters and metallicity, we followed the methodology presented in \citet{Magrini2022A&A...663A.161M} and \citet{Tsantaki+2025}.
The method iteratively combines the results of spectral analysis with photometric and astrometric data, comparing them with stellar evolution models to refine the stellar parameters and thus improve the accuracy and consistency of the results.  The method starts from parameters derived solely from spectroscopy: effective temperature (T$_\mathrm{eff}$), surface gravity (log($g$)), metallicity ([Fe/H]), and microturbulent velocity ($\xi$). Using this initial set, a new log($g$) value is provided based on luminosity and distance data from {\em Gaia}, and the stellar mass is determined via a comparison of the stellar parameters with a grid of Parsec isochrones. The spectroscopic analysis is then repeated with log($g$) fixed to its new value, yielding an updated set of stellar parameters and masses. The procedure is repeated for two iterations, after which convergence is typically achieved, resulting in parameters that show improved agreement with the isochrone grid.
The outcomes of this process are the stellar effective temperature, log($g$), metallicity, $\xi$, mass, radius, and age,  Table~\ref{tab:stellar_par}  summarize our results. The typical precision (1$\sigma$) is 4$\%$ for the mass and 3$\%$ for the radius. 

All components of the binaries differ by less than 1$\sigma$ from the binary mean metallicity. We note that, even if our metallicity estimates are affected by systematic zero-point shifts, the primary quantity of interest is the similarity between the two stars.
Evolutionary models also provide age estimates, and one WB (CD-439363 and CD-439366) is very young, less than 1~Gyr.
CD-439363 is also a fast rotator, with a cross-correlation peak width of 32~km~s$^{-1}$, which confirms its young age.
Since common proper-motion pairs can survive the first hundred million years without being gravitationally bound binaries, we also excluded this WB from the final sample.
All other WBs have age estimates well in excess of 1~Gyr, and the relative ages have a typical uncertainty of 25$\%$.
This uncertainty grows for lower-mass stars, which can formally appear older than the best-estimate age of the Universe (Table~\ref{tab:stellar_par}) because their extended main-sequence lifetimes, together with the high sensitivity and limited variation of the fitting parameters during this evolutionary phase (T$_{\rm eff}$, luminosity, and [Fe/H]), reduce the precision of the derived ages.

In conclusion, of the 26 WBs observed, 12 have highly variable radial velocities, 1 has a {\it Gaia} companion, and 1 is young. That leaves 12 suitable systems, corresponding to 46$\%$ of the initial sample.
The median radial velocity standard deviation of the final sample is 4 $\pm$ 6~ms$^{-1}$.

The effective temperature and gravity were used to compute the radial-velocity blueshift produced by convective motions in the stellar atmosphere.
These corrections were measured by cross-correlating the synthetic 3D atmosphere spectra with the ESPRESSO digital mask \citep{Ludwig+2009, Leao+2019}.
Gravitational redshift was computed from the inferred masses and radii.
The final radial velocity was obtained by subtracting the convective and gravitational shifts from the measured (ESPRESSO) radial velocity: $\mathrm{RV} = \mathrm{RV}_E - \mathrm{RV}_{\rm con} - \mathrm{RV}_{\rm GR}$.

The measured radial velocities (RV$_E$), gravitational redshifts (RV$_{GR}$), and convective velocity shifts (RV$_{con}$) are reported in Table~\ref{tab:stellar_par}.
Following \citet{Leao+2019}, in the analysis we added quadratically to the RV$_E$ variance a systematic uncertainty of 40~m s\,$^{-1}$, which comes from the uncertainty of the zero-point calibration.
Table \ref{tab:distances} reports mean distances from parallaxes, angular separations, masses, projected separations, and velocity differences for the 12 binaries of this study.

\section{Bound Newtonian solutions} 
The geometry and analysis method we followed have been fully detailed in \citet{Saglia+2025}. These authors show how, adding
the third velocity component to the {\it Gaia} projected motions and using the {\it Gaia} parallaxes and the derived masses with their uncertainties, it is possible
to constrain the Keplerian solutions. 

For 3 of the 12 WBs studied, we could not find a viable Keplerian orbit. The total velocity difference for these WBs
(pairs 7, 11, and 12)  is much higher than the maximal Newtonian velocity for bound orbits. This can be seen in  the last column of Table \ref{tab:distances}, which  
gives the ratio between the  total velocity difference between the components and 
$\nu = \sqrt{2GM/r}$, the highest velocity difference allowed by Keplerian orbits. For the three WBs without a viable Keplerian solution, we used $\nu = \sqrt{2GM/s} $. 

Table~\ref{tab:orbits} summarises the best-fit distances and the orbital
parameters for the nine WBs for which a Keplerian solution could be found. 
Out of these nine pairs, three (pairs 4, 8, and 9)  require an eccentricity above 0.9. For pair 1 we 
find a peculiar geometry: despite the very small radial velocity difference, the binary is edge-on -- the two stars are crossing in front of the observer. 
The median ratio between the 3D separation ($r$) and the projected separation ($s$) for the nine WBs with Keplerian orbits is 1.7,  slightly higher than  expected from a uniform angle distribution (1.15). The highest $r/s$ ratio is $\sim$5.4 for the `edge-on' pair (pair 1), confirming the exceptional geometry of this binary. 

\citet{Saglia+2025} find that the orbital solutions in  their sample show an excess of systems with high inclination angles ({\it i}), a larger than expected number of orbits close to periastron ($\phi-\phi_0 \sim $0), and a thermal eccentricity distribution. Adding the nine WBs of this work does not  change these findings. 
\section{Discussion} 

About one-quarter of the analysed systems cannot be reproduced with Keplerian solutions, which makes it necessary to examine the possible causes of this discrepancy. For these WBs, the uncertainties in V$_{tot}$ are less than 10$\%$, and  a few percent  in $\nu$; we therefore rule out the possibility that the observed high velocity differences are caused by observational errors.  
Since we cannot demonstrate that these WBs are in bound orbits,  before invoking  new physics,  we explored alternative explanations and 
the possible causes of this disagreement, which we outline as follows.
\paragraph{Contamination by unseen companions or giant planets:}

Although this  hypothesis is very difficult to rule out, the precise radial velocities and the low RUWE parameter would rule out the presence of unseen stellar companions.
One should, for instance, invoke special geometry, and this might
be the case for pair 7, which contains TYC 5346-457-1. This
star has the highest RUWE, and the radial-velocity difference between the components is $\sim$ 1/8 of the total velocity difference (see Table~\ref{tab:distances}). This might indicate 
a low inclination angle, i.e. an almost face-on orbit. 
A potential companion of TYC 5346-457-1, co-planar with the face-on WB orbit, would explain the high RUWE and would not produce  
radial velocity variations.  This is the only `pair' in our sample that may have a third companion.

Similar considerations apply regarding the possible presence of  massive short-period planets. Of the binaries that cannot be fit with Keplerian solutions, only one star has three ESPRESSO observations, while the other stars have between five and ten. It is unlikely that a massive planet in a close orbit
would not be detected, unless in a very peculiar orbit, while a massive exoplanet in a wide orbit would not increase the radial
velocity difference by a substantial amount (Jupiter produces a radial-velocity variation of only $\sim$13 m\,s$^{-1}$ on the Sun).  
Finally, we compared our catalogue with the updated (September 2025) database of confirmed planet-hosting stars \citep[\hyperlink{www.exoplanets.eu}{www.exoplanets.eu};][]{2011A&A...532A..79S} but found no matches. Considering also the relatively low occurrence of giant planets in short-period orbits, we can reasonably rule this out as a possible explanation.

\paragraph{Common proper motion pairs from clusters and associations:}  We investigated whether any star in our sample is associated
with known open clusters or moving groups. We cross-matched
our catalogue with the list of open clusters and moving group members compiled by 
\citet{2023A&A...673A.114H}, which is based on {\it Gaia} DR3 parallaxes
and proper motions, adopting a cross-match radius of up
to 5 arcseconds. No matches were found.
These WBs are not members of known open clusters or moving groups.

\paragraph{Perturbed, unbound orbits:}
Another hypothesis is that WBs are isolated systems. However, since our WBs are several
billion years old, their orbits have likely been perturbed by encounters with other stars or molecular clouds and by Galactic disk instabilities
\citep{Jiang+2010, Hamilton+2024}. \citet{Hamilton+2024}  provide typical timescales for those interactions as a function of the binary's orbital parameters,  
finding that an orbit semi-major axis of $a\sim10^4$ AU marks the transition
between the timescale dominated by encounters and that dominated by galactic tides. 
The disruption timescale for $a=10^4$ AU and a 2 solar mass system is $\sim$5.7 Gyrs \citep[Eq.\ 18 of][]{Hamilton+2024}. 
 These authors also conclude  that orbital eccentricity does not increase with time,  while  orbital semi-major axes increase with age as an effect of these perturbations. The first conclusion implies that the high eccentricities we now observe were not
smaller in the past; the second implies that we may expect an evolution of semi-major axes with stellar age. 
Finally, \citet{Pegnarrubia2021} and \citet{Hamilton+2024} agree
that binaries with semi-major axes $> $10$^5 $ AU are disrupted in a short time ($\ll 1$\,Gyr). 

\begin{figure}
\centering
 \includegraphics[angle=0,width=7.0cm]{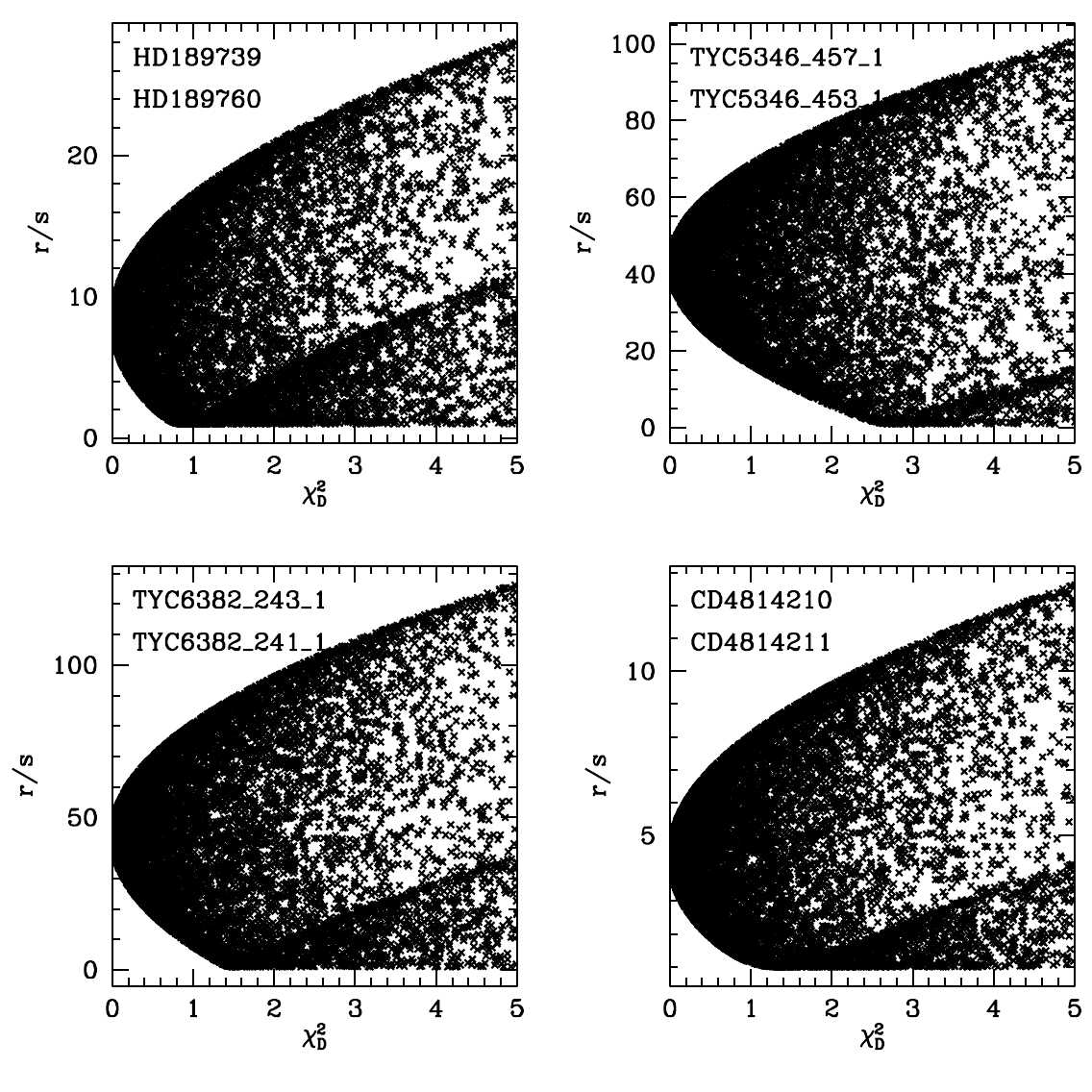}
   \caption{r/s (3D/projected separation) ratio vs distance $\chi ^2$ distribution for the four binaries with no Keplerian solutions: three from this study and one from the HARPS sample (WB24, shown in the top-left panel; \citealt{Saglia+2025}). For all stellar pairs, the most probable r/s are large, from 5 to 50. When multiplying these ratios by the projected separations ($s$), the 3D minimum separation of the three pairs analysed in this paper is  $\sim$ 100 KAU. }
         \label{Fig1}
  \end{figure}
 
We could therefore argue that the WBs with large semi-major axes or large separations are in non-closed orbits, and
that they have been disrupted by encounters or galactic tidal forces.
This hypothesis  seems  reasonable when looking, for instance,  at the \citet{Jiang+2010} simulations (their Fig. 4) that show a secondary peak in the probability density of escaped binaries for systems with a separation of $\sim$ 0.5 Jacobi radii, or  $\sim$ 170 kAU for 2 solar mass stars, independent of the initial stars' separation.  This simulation shows that even when a WB is  disrupted, its  stars can remain in quasi-periodic orbits  within the Jacobi radius for many  gigayears.
This separation is larger than that of most of our systems but is comparable at the order-of-magnitude level for several of our WBs (see Tables~\ref{tab:distances} and \ref{tab:orbits}). Also, \citet{Hamilton+2024} show that binaries with initial separations above a few thousand astronomical units are disrupted in a few gigayears (see their Fig. 6). 
One point that seems at odds with this hypothesis  is that the WBs without a Keplerian solution do not have the largest projected separations ($s$) in our sample (see Table~\ref{tab:distances}: cf.\ 10, 22, and 25 kAU respectively for WBs 7, 11, and 12).  WB24 in \citet{Saglia+2025}, the only one that could not be fitted by a Keplerian solution in that sample, has a projected separation of only 5.77 kAU.  
 However, we cannot exclude the possibility that the 3D separation for these pairs is much larger than the projected one. Indeed, this seems to be the case when looking at the probability distribution of the 3D separation for these binaries. Figure \ref{Fig1} shows the $r/s$ versus distance $\chi ^2$: the most likely 3D separation, $r$, is several times larger than the projected separation, $s$, for all WBs without viable  Keplerian solutions, and larger than $\sim$100 kAU for the three WBs of this study.  Note also that if these stars were bound, such high $r/s$ values would be extremely unlikely. This supports the hypothesis that they are not bound.  We stress that we would miss this information if we used only average distances to the binaries  and projected separations.

Finally, we mention that the orbital eccentricity in WBs has a super-thermal distribution \citep{Hwang+2022}, a feature that cannot be explained by orbital evolution \citep{Hamilton+2024}. While a full discussion on the origin of the distribution is beyond the scope of this work, we note that WBs with high eccentricities can be  produced as the result  of capture \citep{Penarrubia2023} or as chance entrapment in tidal tails \citep{Pegnarrubia2021}. Our  analysis shows that the WBs we studied likely have a common origin, and we verified that they do not belong to known clusters.\ However, we cannot  exclude that  their clusters or associations  of origin  are now dissolved.  
We believe that  our findings add a variable to the eccentricity discussion: since several WBs do not have a Keplerian orbital solution, the super-thermal distribution of eccentricity could have an additional component that has not yet been considered -- the observations are interpreted under the wrong assumption because some of these systems  are  not bound. This could pollute  the very wide systems, but we note that the super-thermal eccentricity is already present at projected separations ($s$) of 1000~AU; at this separation, all the WBs analysed so far can be closely reproduced with Keplerian orbits \citep{Saglia+2025, Chae2025b}.

\section{Conclusions}
By combining precise radial velocities with {\em Gaia} astrometry, even a sample of 12 WBs provides a unique opportunity to study their dynamics. Remarkably, one-third of these systems cannot be reconciled with a Keplerian solution.

Before interpreting these systems as a challenge to Newtonian gravity in the low-acceleration regime, it is essential to explore alternative explanations by improving our understanding of WB evolution and disruption. 
Current models suggest that the three very wide binaries formed in tighter orbits, which subsequently expanded due to Galactic tides and stellar encounters \citep{Hamilton+2024}.
We find that the binaries lacking a Keplerian solution have very large 3D separations, suggesting that they may not be gravitationally bound and could represent disrupted systems.
We emphasise that using individual stellar parallaxes, rather than just the projected separation ($s$) and the system's average distance, is crucial to revealing the large 3D separations of these binaries, which can be hidden by moderate projected distances.

Although the hypothesis of isolated systems can break at any separation, WBs with 3D separations below 10,000 AU may represent the best limit for testing gravity, because 
at this separation,  the binary destruction time is expected to be comparable to or longer than the stellar age \citep{Hamilton+2024}. Ongoing VLT–ESPRESSO observations 
will help address this problem. Low-mass stars, which experience similar accelerations at smaller separations, provide another promising target sample despite being fainter and harder to observe. The upcoming {\em Gaia} DR4, with improved parallaxes and proper motions, will further refine Keplerian orbital constraints.

\begin{acknowledgements}
 The authors thank an anonymous referee for very useful comments and suggestions. 
 Based on data from the European Space Agency (ESA) {\it Gaia} mission, processed by the {\it Gaia} Data Processing and Analysis Consortium (DPAC), https://www.cosmos.esa.int/web/gaia/dpac/consortium.
 MTM acknowledges the support of the Australian Research Council through Future Fellowship grant FT180100194 and through the Australian Research Council Centre of Excellence in Optical Microcombs for Breakthrough Science (project number CE230100006) funded by the Australian Government.
 L.B. and L.M. acknowledge support from INAF through the Large Grants EPOCH and WST, funding for the WEAVE project, the Mini-Grants Checs , the PRIN PNRR Project Cosmic POT (2022X4TM3H, MUR) and from the  HORIZON-INFRA-2024-DEV-01-01 (Grant No. 101183153). J.C. acknowledges support from ANID via Proyecto Fondecyt Regular 1231345, and BASAL project FB210003.
\end{acknowledgements}

\bibliographystyle{aa} 
\bibliography{wb4.bib} 

\begin{appendix}
\section{Observed systems not suitable for our test}
\label{appendix}
\begin{itemize}\setlength\itemsep{0pt}
\item CD-28476 and CPD-28132 : CD-28476 has a RV variability of 160 ms$^{-1}$ over 5 observations.
\item TYC8862-728-1 and TYC8862-1029-1: TYC8862-1029-1 has only 16 ms$^{-1}$ RV rms, but the observations in the the second year cluster around a value about 30 ms$^{-1}$ lower than the observations of the first year. We consider this star a likely long-period binary.
\item TYC-6462-304-1 and TYC6462-326-1: The RV TYC-6462-304-1 varies between 23 and 38 km$^{-1}$.
\item TYC7080-660-1 and TYC7080-472-1: TYC7080-472-1 shows an RMS of 15 kms$^{-1}$.
\item HD57376 and HD57257: HD57257 shows substantial variability (100 ms$^{-1}$) for observations taken 10 days apart.
\item CD-68467 and CD-68468: CD-68468 shows 287 ms$^{-1}$ radial velocity rms. 
\item TYC7745-788-1 and TYC7745-1033-1: TYC7745-788-1 shows 127 ms$^{-1}$ spread, with a large difference between one year and next year.
\item CD-456209 and HD296796: both stars show large radial velocity variability, probably a quadruple system.
\item TYC7644-884-1 and TYC7644-399-1: TYC7644-884-1 shows a variability of 140 ms$^{-1}$: observations taken 1 year later differ by 300 ms$^{-1}$ 
\item TYC565-731-1 the RV between the two observations varies by 85 ms$^{-1}$, with an associated uncertainty of 2 ms$^{-1}$. The star is
likely double.
\item TYC8445-1018-1 in four observations within 16 days the radial velocity varies between 993 and 1083 ms$^{-1}$, but the observations
have large associated uncertainties (8-14 ms$^{-1}$). Our chemical analysis, in addition, shows a difference of 0.5 dex in [Fe/H] with respect to its companion TYC 8445-1073-1. 
\item TYC9059-134-1 has an rms of 36 ms$^{-1}$ on 3 observations within a few weeks, each with an associated uncertainty of 9 ms$^{-1}$. Its companion, TYC9059-376-1 has only one observation.
\item BD-20655 and BD-20658: the {\it Gaia} DR3 catalogue shows that BD-20655 has a close, faint companion with same parallax and proper motion. 
\end{itemize} 

\section{Tables} 
\onecolumn
\begin{table}[ht]
\caption{Gaia DR3 data for the final sample stars.}
\label{table1}
\centering
\small
\begin{tabular}{l l c c c c c c}
\hline\hline
No. & Name & $\alpha$ & $\delta$ & Parallax & $\mu_{\alpha}$ & $\mu_{\delta}$ & RUWE \\ 
& & (deg) & (deg) & (mas) & (mas/yr) & (mas/yr) & \\ 
\hline\hline
 1 & TYC 7527-458-1 & 6.54649 & -40.97890 & $1.919 \pm 0.016$ & $11.021 \pm 0.014$ & $-8.974 \pm 0.017$ & 1.17 \\ 1 & TYC 7527-959-1 & 6.55700 & -40.98052 & $1.905 \pm 0.018$ & $10.970 \pm 0.016$ & $-8.956 \pm 0.019$ & 1.23 \\ 2 & TYC 8040-427-1 & 25.78047 & -46.38649 & $3.781 \pm 0.014$ & $3.186 \pm 0.011$ & $12.973 \pm 0.013$ & 0.98 \\ 2 & TYC 8040-648-1 & 25.79858 & -46.39370 & $3.746 \pm 0.014$ & $3.141 \pm 0.011$ & $12.961 \pm 0.014$ & 1.02 \\ 3 & TYC 5896-1344-1 & 65.37377 & -20.39827 & $4.801 \pm 0.015$ & $9.971 \pm 0.013$ & $-12.399 \pm 0.015$ & 0.99 \\ 3 & TYC 5896-971-1 & 65.43523 & -20.44224 & $4.800 \pm 0.019$ & $9.942 \pm 0.019$ & $-12.337 \pm 0.020$ & 1.26 \\ 4 & BD+05 650 & 66.46340 & 6.07853 & $10.305 \pm 0.017$ & $29.119 \pm 0.017$ & $-8.078 \pm 0.013$ & 0.94 \\ 4 & BD+05 653 & 66.53689 & 6.05450 & $10.354 \pm 0.017$ & $28.705 \pm 0.017$ & $-8.318 \pm 0.012$ & 0.93 \\ 5 & TYC 9492-248-1 & 79.45019 & -82.53983 & $2.448 \pm 0.010$ & $-8.528 \pm 0.012$ & $21.206 \pm 0.012$ & 1.02 \\ 5 & TYC 9492-241-1 & 79.52725 & -82.55538 & $2.455 \pm 0.010$ & $-8.656 \pm 0.012$ & $21.246 \pm 0.012$ & 1.06 \\ 6 & HD 294014 & 80.54664 & -3.46269 & $3.191 \pm 0.010$ & $-25.726 \pm 0.017$ & $16.973 \pm 0.013$ & 1.13 \\ 6 & HD 294012 & 80.57905 & -3.38525 & $3.162 \pm 0.019$ & $-25.700 \pm 0.018$ & $17.009 \pm 0.013$ & 1.14 \\ 7& TYC 5346-457-1 & 84.94444 & -7.84519 & $4.671 \pm 0.023$ & $24.642 \pm 0.021$ & $-38.050 \pm 0.019$ & 1.45 \\ 7& TYC 5346-453-1 & 84.95605 & -7.83808 & $4.625 \pm 0.016$ & $23.814 \pm 0.014$ & $-38.554 \pm 0.012$ & 1.14 \\ 8& TYC 7068-419-1 & 85.26147 & -35.69142 & $2.506 \pm 0.011$ & $7.694 \pm 0.012$ & $1.676 \pm 0.013$ & 0.96 \\ 8 & TYC 7068-433-1 & 85.27586 & -35.68482 & $2.521 \pm 0.010$ & $7.734 \pm 0.011$ & $1.732 \pm 0.012$ & 0.90 \\ 9 & TYC 752-1389-1 & 104.52843 & 10.64266 & $12.818 \pm 0.014$ & $-25.424 \pm 0.015$ & $-20.182 \pm 0.013$ & 0.94 \\ 9 & TYC 752-1649-1 & 104.54791 & 10.60863 & $12.814 \pm 0.013$ & $-25.275 \pm 0.015$ & $-20.740 \pm 0.013$ & 0.97 \\ 10 & HD 94255 & 163.09504 & -23.20317 & $11.383 \pm 0.019$ & $10.920 \pm 0.018$ & $5.787 \pm 0.019$ & 1.05 \\ 10 & HD 94271 & 163.13797 & -23.18022 & $11.386 \pm 0.017$ & $11.535 \pm 0.016$ & $6.396 \pm 0.017$ & 0.97 \\ 11 & TYC 6382-243-1 & 328.38811 & -20.17357 & $2.712 \pm 0.019$ & $-24.635 \pm 0.019$ & $-33.375 \pm 0.018$ & 0.77 \\ 11 & TYC 6382-241-1 & 328.40533 & -20.17245 & $2.679 \pm 0.020$ & $-24.717 \pm 0.020$ & $-33.577 \pm 0.020$ & 0.87 \\ 12 & CD-4814210 & 336.69311 & -48.35999 & $6.951 \pm 0.017$ & $120.005 \pm 0.014$ & $-17.398 \pm 0.014$ & 1.04 \\ 12 & CD-4814211 & 336.74459 & -48.32539 & $6.976 \pm 0.014$ & $119.710 \pm 0.011$ & $-17.083 \pm 0.012$ & 0.92 \\ \hline 
\hline
\end{tabular}
\tablefoot{First column indicates the pairs' number, as used in the text.}
\end{table}

\begin{table}[ht]
\caption{Derived stellar parameters and radial velocities.}
\label{tab:stellar_par}
\centering
\small
\setlength{\tabcolsep}{2pt}
\begin{tabular}{@{} l l l l c c c c c c c l @{}}
\hline\hline\hline\hline
No & Star & Age & T$_{\rm eff}$ & $\log g$ & [Fe/H] & Mass & Radius & RV$_{\rm gr}$ & RV$_{\rm con}$ & Mask/N & RV$_E$ \\
   &      & (Gyr) & (K)           &          &        & (M$_\odot$) & (R$_\odot$) & (km~s$^{-1}$) & (km~s$^{-1}$) &        & (km~s$^{-1}$) \\
1  &  TYC 7527-458-1  &  5.0   &   6368 $\pm$ 30   &  3.97 $\pm$ 0.12  &  -0.45 $\pm$ 0.08  &  1.04   &        1.74     &      0.38 &  -0.90 &  F9/6 &  -9.116  $\pm$ 0.003\\        
1  &  TYC 7527-959-1  &  2.3   &   6807 $\pm$ 42   &  4.20 $\pm$ 0.04  &  -0.38 $\pm$ 0.08  &  1.21   &        1.45     &      0.53 &  -0.91 &  F9/4 &  -8.980  $\pm$ 0.021\\      
2  &  TYC 8040-427-1  &  7.2   &   6034 $\pm$ 100  &  4.14 $\pm$ 0.11  &  -0.33 $\pm$ 0.12  &  0.99   &        1.40     &      0.45 &  -0.74 &  F9/7 &  32.957  $\pm$ 0.005\\      
2  &  TYC 8040-648-1  &  9.2   &   5840 $\pm$ 96   &  4.17 $\pm$ 0.11  &  -0.38 $\pm$ 0.12  &  1.04   &        1.40     &      0.47 &  -0.69 &  F9/6 &  32.826  $\pm$ 0.007\\      
3  &  TYC 5896-1344-1 &  6.0   &   5830 $\pm$ 31   &  4.32 $\pm$ 0.04  &  -0.10 $\pm$ 0.10  &  0.96   &        1.12     &      0.54 &  -0.63 &  F9/6 &  25.552  $\pm$ 0.014\\      
3  &  TYC 5896-971-1  &  7.9   &   5790 $\pm$ 67   &  4.40 $\pm$ 0.07  &  -0.04 $\pm$ 0.10  &  0.97   &        1.02     &      0.60 &  -0.63 &  F9/7 &  25.630  $\pm$ 0.012\\      
4  &  BD+05 650       &  6.7   &   6032 $\pm$ 70   &  4.30 $\pm$ 0.06  &  -0.21 $\pm$ 0.11  &  1.00   &        1.17     &      0.54 &  -0.71 &  F9/8 &  23.286  $\pm$ 0.010\\      
4  &  BD+05 653       &  6.1   &   5856 $\pm$ 55   &  4.37 $\pm$ 0.03  &  -0.09 $\pm$ 0.10  &  0.97   &        1.07     &      0.57 &  -0.64 &  F9/9 &  23.337  $\pm$ 0.022\\      
5  &  TYC 9492-248-1  &  15.7  &   5764 $\pm$ 76   &  4.05 $\pm$ 0.13  &  -0.90 $\pm$ 0.10  &  0.75   &        1.36     &      0.35 &  -0.58 &  G2/3 &  18.534  $\pm$ 0.001\\      
5  &  TYC 9492-241-1  &  16.9  &   5828 $\pm$ 113  &  4.14 $\pm$ 0.09  &  -0.81 $\pm$ 0.10  &  0.76   &        1.23     &      0.39 &  -0.58 &  G2/3 &  18.467  $\pm$ 0.002\\      
6  &  HD 294014       &  10.6  &   5758 $\pm$ 87   &  4.07 $\pm$ 0.10  &  -0.64 $\pm$ 0.11  &  0.94   &        1.55     &      0.38 &  -0.67 &  F9/5 &  23.755  $\pm$ 0.002\\      
6  &  HD 294012       &  10.3  &   5800 $\pm$ 73   &  4.01 $\pm$ 0.11  &  -0.62 $\pm$ 0.10  &  0.94   &        1.58     &      0.38 &  -0.70 &  F9/5 &  23.819  $\pm$ 0.004\\      
7  &  TYC 5346-457-1  &  8.6   &   5820 $\pm$ 95   &  4.45 $\pm$ 0.12  &  -0.49 $\pm$ 0.11  &  0.82   &        0.89     &      0.58 &  -0.61 &  G8/7 &  -67.23  $\pm$ 0.001\\      
7  &  TYC 5346-453-1  &  10.8  &   5726 $\pm$ 79   &  4.47 $\pm$ 0.11  &  -0.60 $\pm$ 0.10  &  0.75   &        0.84     &      0.57 &  -0.58 &  G8/7 &  -67.09  $\pm$ 0.001\\      
8  &  TYC 7068-419-1  &  3.2   &   6136 $\pm$ 56   &  4.20 $\pm$ 0.06  &  0.17  $\pm$ 0.12  &  1.24   &        1.46     &      0.54 &  -0.76 &  F9/8 &  5.403   $\pm$ 0.011\\      
8  &  TYC 7068-433-1  &  5.4   &   5922 $\pm$ 57   &  4.17 $\pm$ 0.01  &  0.14  $\pm$ 0.10  &  1.11   &        1.44     &      0.49 &  -0.70 &  F9/7 &  5.198   $\pm$ 0.012\\      
9  &  TYC 752-1389-1  &  10.8  &   5338 $\pm$ 41   &  4.51 $\pm$ 0.01  &  -0.42 $\pm$ 0.09  &  0.73   &        0.78     &      0.59 &  -0.47 &  G2/7 &  -17.41  $\pm$ 0.004\\      
9  &  TYC 752-1649-1  &  15.1  &   5304 $\pm$ 39   &  4.58 $\pm$ 0.09  &  -0.35 $\pm$ 0.09  &  0.81   &        0.76     &      0.68 &  -0.44 &  G8/8 &  -17.37  $\pm$ 0.001\\      
10 &  HD 94255         &  3.7   &   6046 $\pm$ 74   &  4.36 $\pm$ 0.09  &  -0.02 $\pm$ 0.11  &  1.07   &        1.14     &      0.59 &  -0.69 &  F9/8 &  12.117  $\pm$ 0.003\\      
10 &  HD 94271         &  3.5   &   6044 $\pm$ 78   &  4.37 $\pm$ 0.09  &  0.01  $\pm$ 0.10  &  1.08   &        1.13     &      0.61 &  -0.69 &  F9/8 &  12.281  $\pm$ 0.005\\      
11 &  TYC 6382-243-1  &  7.7   &   6002 $\pm$ 85   &  3.99 $\pm$ 0.10  &  -0.46 $\pm$ 0.10  &  1.03   &        1.69     &      0.38 &  -0.76 &  F9/5 &  16.659  $\pm$ 0.004\\      
11 &  TYC 6382-241-1  &  10.5  &   5790 $\pm$ 71   &  3.99 $\pm$ 0.11  &  -0.64 $\pm$ 0.12  &  0.97   &        1.64     &      0.37 &  -0.71 &  F9/3 &  16.296  $\pm$ 0.004\\    
12 &  CD-4814210      &  7.5   &   5992 $\pm$ 48   &  4.20 $\pm$ 0.20  &  -0.21 $\pm$ 0.10  &  0.99   &        1.26     &      0.49 &  -0.58 &  G2/9 &  -66.14  $\pm$ 0.003\\      
12 &  CD-4814211      &  6.1   &   5978 $\pm$ 38   &  4.17 $\pm$ 0.20  &  -0.11 $\pm$ 0.11  &  1.01   &        1.16     &      0.55 &  -0.62 &  G2/8 &  -66.650 $\pm$ 0.001\\
\hline
\end{tabular}
\tablefoot{RV$_E$ is the radial velocity as measured by ESPRESSO (cfr. Eq. 1).}
\end{table}

\onecolumn
\begin{longtable}{l c c c c c c c c c c} 
\caption{Mean distance, angular separation, mass, projected separation, and velocities for the WB.} 
\label{tab:distances} \\
\hline\hline
No &  MeanD & $\theta$ & Mass & $s$   & $V_{\rm ra}$        &  $V_{\rm dec}$      & $V_{\rm rad}$       & $V_{\rm tot}$     & $\nu$            & $V_{\rm tot}/\nu$\\
   &  (pc)  & (") & $(M_\odot)$   & (AU)       & (m~s$^{-1}$)       &  (m~s$^{-1}$)      &  (m~s$^{-1}$)     & (m~s$^{-1}$) & (m~s$^{-1}$)\\  
\hline
\endfirsthead

\caption{continued.}\\
\hline\hline
No &  MeanD & $\theta$ & Mass & $s$   & $V_{\rm ra}$        &  $V_{\rm dec}$      & $V_{\rm rad}$       & $V_{\rm tot}$     & $\nu$            & $V_{\rm tot}/\nu$\\
\hline
\endhead
\hline
\endfoot

\hline
\multicolumn{11}{p{\textwidth}}{\rule{0pt}{3ex}\small \textbf{Notes.} $V_{\text{tot}}/\nu$ is the ratio between the observed velocity and the maximum Newtonian velocity. Transverse velocities are computed using the distances of Table \ref{tab:orbits}, except for binaries 7, 11, 12, where no orbit is found and the mean distance is used. For these three WBs $\nu$ is computed using the projected separation {\it s} instead of the 3D separation {\it r}.} \\
\endlastfoot

1  &  523.0 &  29.2  &  2.3 & 15067 & -145.2 $\pm$  10.9 &   60.6 $\pm$ 13.1 &    2.4 $\pm$ 45.3 & 157.4 $\pm$ 15.7 & 224.3 $\pm$  2.5 &  0.7 $\pm$ 0.10\\
2  &  265.7 &  51.9  &  2.0 & 13782 & -59.1  $\pm$  4.2  &  -24.5 $\pm$  5.2 & -208.0 $\pm$ 40.9 & 217.7 $\pm$ 35.9 & 326.4 $\pm$  7.3 &  0.7 $\pm$ 0.13\\
3  &  208.3 & 260.9  &  1.9 & 54249 & -57.7  $\pm$  4.7  &   98.0 $\pm$  5.2 &   13.7 $\pm$ 44.0 & 114.6 $\pm$ 10.7 & 153.0 $\pm$  2.1 &  0.7 $\pm$ 0.41\\
4  &  96.8  & 276.9  &  2.0 & 26789 & -214.7 $\pm$  2.3  & -102.9 $\pm$  1.7 &  -47.5 $\pm$ 46.7 & 242.8 $\pm$ 11.0 & 274.6 $\pm$  2.6 &  0.9 $\pm$ 0.04\\
5  &  407.9 &  66.5  &  1.5 & 27096 & -245.6 $\pm$  7.0  &   74.8 $\pm$  6.8 & -111.2 $\pm$ 40.1 & 279.8 $\pm$ 18.2 & 309.5 $\pm$  5.4 &  0.9 $\pm$ 0.06\\
6  &  314.8 & 302.2  &  1.9 & 94840 &  45.0  $\pm$  7.7  &   49.1 $\pm$  5.9 &  101.1 $\pm$ 40.2 & 121.1 $\pm$ 33.7 & 186.1 $\pm$  2.9 &  0.7 $\pm$ 0.17\\
7  &  215.1 &  48.7  &  1.6 & 10475 &  844.5 $\pm$  25.5 &  514.5 $\pm$ 22.6 & -124.2 $\pm$ 41.2 & 996.6 $\pm$ 26.4 & 515.6 $\pm 11.1$ &  1.9 $\pm$ 0.10\\
8  &  397.8 &  48.3  &  2.4 & 19202 &  68.8  $\pm$  6.3  &  105.6 $\pm$  7.0 & -212.3 $\pm$ 43.2 & 246.9 $\pm$ 36.2 & 317.1 $\pm$  1.9 &  0.8 $\pm$ 0.11\\
9  &  78.0  & 140.9  &  1.5 & 10965 &  60.8  $\pm$  1.6  & -201.8 $\pm$  1.4 &  -61.8 $\pm$ 40.2 & 219.6 $\pm$ 12.3 & 433.6 $\pm$  0.9 &  0.5 $\pm$ 0.02\\
10 &  87.8  & 164.4  &  2.1 & 14434 & 255.1  $\pm$  2.2  &  253.0 $\pm$  2.2 &  149.4 $\pm$ 40.4 & 389.1 $\pm$ 16.7 & 506.7 $\pm$  9.8 &  0.8 $\pm$ 0.04\\
11 & 371.0  &  58.5  &  2.0 & 21693 & 144.1  $\pm$  48.6 &  354.9 $\pm$ 47.9 &  402.0 $\pm$ 40.4 & 555.3 $\pm$ 38.0 & 404.3 $\pm$  5.0 &  1.4 $\pm$ 0.10\\
12 & 143.6  & 175.2  &  2.1 & 25157 & 201.1  $\pm$  11.9 & -214.5 $\pm$ 12.4 &  530.0 $\pm$ 40.1 & 606.0 $\pm$ 37.9 & 382.0 $\pm$  5.8 &  1.6 $\pm$ 0.10\\
\end{longtable}
  
 \begin{table*}
 \caption{Orbital parameters for the Newtonian solutions of the nine WBs that could be fit.} 
 \centering 
 \begin{tabular}{l c c c c c c c c c c}
   \hline
No & D$_1$  & D$_2$ & $i$       &  $e$      & $\phi-\phi_0$ & $r$       & $a$            & $\chi^2$ &  log $P$  & log $A_N$ \\
   & (pc)   & (pc)  & (deg)     &           & (deg)         & (AU)      & (AU)           &         &    (yrs)  &  (m~s$^{-2}$) \\
\hline
1  &  517.2 & 516.8 & 87.9 &  0.23 & 286.9 & 80931 & 79677  & 3.26 &  7.172 & -11.689\\ 
2  &  265.7 & 265.5 & 81.0 &  0.81 & 148.5 &  34810 & 31348  & 3.62 &  6.585 & -11.041\\ 
3  &  208.6 & 208.0 & 81.1 &  0.56 & 293.1 & 137758 & 156677 & 1.30 &  7.663 & -12.196\\ 
4  &   96.9 &  96.7 & 66.0 &  0.96 & 330.3 &  46170 & 105859 & 1.62 &  7.391 & -11.069\\ 
5  &  407.4 & 407.3 & 42.3 &  0.80 & 272.4 &  27584 & 75439  & 0.40 &  7.230 & -10.722\\ 
6  &  313.5 & 313.9 & 78.2 &  0.44 & 133.2 &  95468 & 82805  & 1.86 &  7.242 & -11.988\\ 
8  &  397.4 & 397.3 & 62.2 &  0.97 & 163.5 &  41220 & 523901 & 1.01 &  6.895 & -11.005\\ 
9  &  78.0  & 78.0  & 69.8 &  0.92 & 166.8 &  14567 & 9797   & 0.45 &  5.893 & -10.655\\ 
10 &  87.8  & 87.8  & 69.8 &  0.77 & 130.8 &  14968 & 18238  & 0.01 &  6.224 & -10.170\\ 
   \hline 
   \end{tabular} 
 \label{tab:orbits}
 \end{table*} 
\end{appendix}
\end{document}